\documentclass[aps,prd,eqsecnum,amsmath,onecolumn,notitlepage,nofootinbib,superscriptaddress,10pt]{webofc}
\usepackage[varg]{txfonts}   
\usepackage{custom}

\begin{document}

\title{Relativistic fluctuations in stochastic fluid dynamics}

\author{\firstname{Xin} \lastname{An}\inst{1}\fnsep\thanks{\email{xin.an@ncbj.gov.pl}} \and
        \firstname{Gökçe} \lastname{Başar}\inst{2}\fnsep\thanks{\email{gbasar@unc.edu}} \and
        \firstname{Mikhail} \lastname{Stephanov}\inst{3}\fnsep\thanks{\email{misha@uic.edu}} \and
        \firstname{Ho-Ung} \lastname{Yee}\inst{3}\fnsep\thanks{\email{hyee@uic.edu}}
}

\institute{National Centre for Nuclear Research, 02-093 Warsaw, Poland 
\and       Department of Physics and Astronomy, University of North Carolina, Chapel Hill, North Carolina 27599, USA 
\and       Department of Physics, University of Illinois, Chicago, Illinois 60607, USA
}

\abstract{
  The state-of-the-art theoretical formalism for a covariant description of non-Gaussian fluctuation dynamics in relativistic fluids is discussed.
}

\maketitle

\section{Introduction}
\label{sec:intro}

Fluctuating hydrodynamics is a powerful tool for exploring complex and critical phenomena in non-equilibrium systems that possess a small number of degrees of freedom. Such scenarios are realized in relativistic heavy-ion collisions where a few thousand particles are produced, especially when the hypothetical QCD critical point is approached. Therefore, fully establishing the framework for relativistic fluctuating hydrodynamics is necessary for interpreting observables in experiments that are sensitive to fluctuations and criticality. Developing a covariant description of non-Gaussian fluctuation dynamics in stochastic fluids is a crucial step toward achieving this ambitious goal.

In this proceeding, we will briefly formulate the essential theoretical development of deterministic fluctuating hydrodynamics from a general perspective. More specifically, we will begin by reviewing the results of Ref.~\cite{An_2023} in Sec.~\ref{sec:equations} and \ref{sec:conflent}, where we presented the covariant formalism for the non-equilibrium evolution of non-Gaussian fluctuations in relativistic fluids. It on one hand follows the approach used in earlier work ~\cite{An:2019rhf} and \cite{An:2019fdc}, where the covariant dynamical description was developed but only for Gaussian fluctuations, and on the other hand extends the subsequent work~\cite{An_2021}, where a generic formalism for non-Gaussian fluctuation dynamics was established, yet it has not been implemented covariantly. In Sec.~\ref{sec:RPA}, we will illustrate how the number of independent $n$-point correlation functions can be significantly reduced, upon the use of certain approximation which facilitates easier numerical implementation. We shall keep the formulation as general as possible, allowing it to be applied to arbitrary $n$ and regimes not relying on the separation of relaxation time scales.

\section{Theoretical framework}

\subsection{Fluctuation evolution equations}
\label{sec:equations}

We start from the {\em covariant} Langevin equation for a set of stochastic
fields $\psi_i$ (such as conserved quantities including charge density and energy-momentum density) where the subscript $i$ labels different fields: 
\begin{align}\label{eq:dtpsi}
  u\cdot\partial\psi_i=F_i+\xi_i\,,
\end{align}
where $F_i$ is the drift force, $\xi_i=H_{ij}\eta_j$ is the multiplicative noise whose amplitude $H_{ij}$ is related to the Onsager matrix via $Q_{ij}\equiv H_{ik}H_{kj}$, with a Gaussian form $\av{\eta_{i}(x_1)\eta_{j}(x_2)}=2\delta_{ij}\delta^{(4)}(x_1-x_2)$. One should keep in mind that the velocity $u\equiv u(\psi_i)$ may or may not be chosen as one of the independent variables in $\psi_i$.

The importance of fluctuations can be studied by the multi-point correlation functions. The ``raw'' $n$-point correlation functions are defined by $G_n\equiv G_{i_1\dots i_n}\equiv\av{\phi_{i_1}(x_1)\dots\phi_{i_n}(x_n)}$ where $\phi=\psi-\av{\psi}$. The connected correlation functions $G^{\rm c}_n\equiv G^{\rm c}_{i_1\dots i_n}$, which are directly related to the experimental observables (such as cumulants of particle multiplicity distribution), can be obtained via the relation
\begin{equation}\label{eq:Gn_c-Gn}
  G^{\rm c}_{i_1\dots i_n}=\sum_{k=1}^n\sum_{\{n_1,\dots,n_k\}}(-1)^{k-1}(k-1)!\nf G_{\underbrace{i_1\dots i_{n_1}}_{n_1}}G_{\underbrace{i_{n_1+1}\dots i_{n_1+n_2}}_{n_2}}\dots G_{\underbrace{i_{n-n_k+1}\dots i_{n}}_{n_k}}\bigg|_{\overline{1\dots n}}\,,
\end{equation}
where $\nf \equiv n!/n_1!\dots n_k!\,k_1!\dots k_n!$, the inner sum is over all ordered sets of integer numbers
$\{n_1,\dots,n_k\}$, such that $n_1\leqslant n_2\leqslant\dots\leqslant n_k$ and $n_1+\dots+n_k=n$. Each set
describes a partition of the $n$ indices $i_1,\dots,i_n$ into $k$ groups in a way that each term in the sum in Eq.~\eqref{eq:Gn_c-Gn} is different.

\begin{figure}[ht]
  \centering  
  \includegraphics[scale=.77]{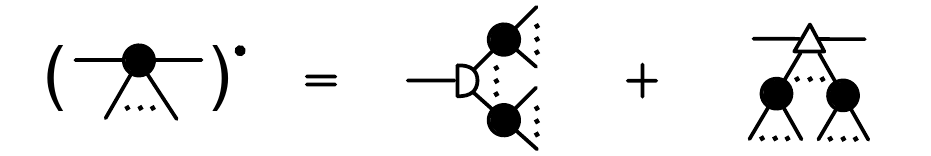}
  \caption{Diagrammatic representation of the evolution equations for generic
    multipoint connected correlation functions, with all possible combinatorial arrangements at tree level.}
  \label{fig:diagrams-n}
\end{figure}

Using Eqs.~\eqref{eq:dtpsi} and \eqref{eq:Gn_c-Gn}, it shall be straightforward to derive the evolution equation for $G_n^{\rm c}$. However, this system of equations is rather complicated and not in a closed form. Following Ref.~\cite{An_2021} we expand and truncate these equations in two independent small parameters, $\varepsilon$ and $\varepsilon_q$, controlling the loop and gradient expansion respectively. The resulting power counting is specified as
\begin{equation}\label{eq:powercounting}
  G^c_{i_1\dots i_n}\sim\varepsilon^{n-1}\,,
  \quad L_{i_1,\,i_2\dots i_n}\sim \varepsilon_q + \mathcal O(\varepsilon_q^2)\,, \quad
  Q_{i_1i_2,\,j_1\dots j_m}\sim\varepsilon_q^2\,\varepsilon\,, \quad
  u\cdot\partial\sim\varepsilon_q^2\,,
\end{equation}
where indices following comma ``\,,\,'' denote the derivative with respect to the fields $\psi$. With the help of Eq.~\eqref{eq:powercounting}, one finds the generic evolution equations for $n$-point connected correlation function at leading order $\varepsilon_q^2\,\varepsilon^{n-1}$:
\begin{multline}\label{eq:dtGn_c_general}
  \,u\cdot\partial^{(x)} G^{\rm c}_{i_1\dots i_n}(x_1,\dots, x_n)=n\Big[-\left(y_1\cdot\partial u\right)\cdot\frac{\partial}{\partial x_1}G^{\rm c}_{i_1\dots i_n}\\
  +\sum_{k=1}^{n-1}\sum_{\substack{\{n_1,\dots,n_k\}\\n_1+\dots+n_k=n-1}}\nf L_{i_1,\,j_1\dots j_k}G^{\rm c}_{j_1\underbrace{i_2\dots}_{n_1}}\dots G^{\rm c}_{j_k\underbrace{\dots i_n}_{n_k}}\\
  +(n-1)\sum_{k=0}^{n-2}\sum_{\substack{\{n_1,\dots,n_k\}\\n_1+\dots+n_k=n-2}}\nf Q_{i_1i_2,\,j_1\dots j_k}G^{\rm c}_{j_1\underbrace{i_3\dots}_{n_1}}\dots G^{\rm c}_{j_k\underbrace{\dots i_n}_{n_k}}\Big]_{\overline{1\dots n}}\,,
\end{multline}
where the time derivative is taken in the rest frame of the fluid at midpoint $x\equiv \sum_{i=1}^n x_i/n$, $y_i\equiv x_i-x$ is the separation vector, $[\dots]_{\overline{1\dots n}}$ is the ``averaged'' permutation over indices labeled by $i$, and
\begin{equation}\label{eq:L}
  L_{i,\,j_1j_2\dots j_n}=F_{i,\,j_1j_2\dots j_n}-u^\mu_{,\,j_1\dots j_n}(\partial_\mu\psi_i)-\left[n\delta_{ij_1}u^\mu_{,\,j_2\dots j_{n}}\partial_\mu^{(j_1)}\right]_{\overline{1\dots n}}
\end{equation}
is a multilinear operator where the last two terms are consequence of $u=u(\psi_i)$. The spacetime arguments for $L$, $Q$ and $G^{\rm c}$, associated with the indices of those quantities, are suppressed on the right hand side. Eq.~\eqref{eq:dtGn_c_general} can be more straightforwardly represented in diagrams in Fig.~\ref{fig:diagrams-n}. These diagrams can be categorized into two groups, one from drift force and the other from the noise, corresponding to the second and third line in Eq.~\eqref{eq:dtGn_c_general} respectively.

\subsection{Confluent formalism}
\label{sec:conflent}

In Sec.~\ref{sec:equations} we have adopted an alternative strategy presented in Refs.~\cite{An:2019rhf} and \cite{An:2019fdc} but generalize it to non-Gaussian fluctuations. This approach is more suitable to derive the Lorentz covariant fluctuation evolution equations. In addition to rewriting the equation in a covariant form with quantities measured in the local rest frame (like how it shall be done for a one-point function), we also need to implement it for multi-point equal-time functions where the confluent formalism is necessary to covariantly describe fluctuating fields at different points.

In the confluent formalism we first introduce the confluent $n$-point connected function
\begin{equation}\label{eq:bar-G}
  \bar G^{\rm c}_n\equiv\bar G^{\rm c}_{i_1\dots i_n}(x_1,\dots,x_n)\equiv\Lambda_{i_1}^{~j_1}(x_1-x)\ldots\Lambda_{i_n}^{~j_n}(x_n-x) G^{\rm c}_{j_1\dots j_n}(x_1,\dots,x_n)\,,
\end{equation}
where $\Lambda(x_i-x)$ is a Lorentz boost defined by $\Lambda(\Delta x)u(x+\Delta x)=u(x)$. In Eq.~\eqref{eq:bar-G} fluctuations at different points are boosted into the same frame (chosen as the local rest frame at the midpoint $x$). The separation four-vector $y_i(x)\equiv x_i-x$ can be expressed in terms of the local tetrad consisting of vector $u(x)$ and triad $e_a(x)$ with $a,b=1,2,3$: $y(x)=e_a(x)y^a$ where we have imposed the equal-time constraint $u\cdot y=0$. The generalized multi-point Wigner transform is then defined as
\begin{align}\label{eq:W-ya}
  W_n(x;\bm q_1,\dots, \bm q_n)=\bigintsss \left[\prod_{i=1}^n
  d^3y_i^a\,e^{-iq_{ia}y_i^a}\right]\delta^{(3)}\left(\frac{1}{n}\sum_{i=1}^ny_i^a\right)\bar
  G_n^\tc(x+e_a y_1^a,\ldots,x+e_a y_n^a)\,,
\end{align}
where ${\bm q}=\{q^a\}$ is the wavenumber conjugate to $y^a$. This definition eliminates the dependence on $x$ of the transform kernel, making it convenient for a covariant theory. It is easy to check that the derivative $\dyafxd$ commutes with the Wigner transform, and $\partial_a^{(y_i)}\to i q_{ia}$.

As the midpoint moves, only the changes respect the local rest frame of the midpoint are measured, not the changes resulted from the change of frame itself, we thus boost $G_n^{\rm c}$ to the frame before the movement to eliminate the kinematic change, at the same time keep the equal-time constraint to preserve the relative positions of the $n$ different points. Taking this two issues into account the confluent derivative is given by
\begin{equation}
    \label{eq:Wrcfd-Wcfd}
    \cfd_\mu W_{i_1\dots i_n} \equiv \dyafxd_\mu W_{i_1\dots i_n}
    - n\left( \bar\omega_{\mu i_1}^{\jj } W_{\jj i_2\dots i_n}
    - \econ_{\mu b}^{a}q_{1a}\partial^{(q_1)}_b
      W_{i_1\dots i_n}
    \right)_{\overline{1\dots n}}\,,
  \end{equation}
where $\bar\omega_{\mu i}^j=u_{i} \partial_\mu u^{j}-u^{j} \partial_\mu u_{i}$ and $\econ_{\mu b}^a=e_\nu^a\partial_\mu e^\nu_b$ are connections, $\dyafxd$ is a derivative with respect to $x$ with $q_a$ fixed, while $\partial^{(q)}_a$ is a derivative with respect to $q_a$ with $x$ fixed.

The evolution equation of Wigner functions can be obtained by projecting Eq.~\eqref{eq:Wrcfd-Wcfd} along
$u(x)$ and using Eqs.~\eqref{eq:dtGn_c_general} and \eqref{eq:W-ya}:
\begin{align}\label{eq:cfd-W_c}
  &u\cdot\cfd W_{i_1\dots i_n}(x;\bm q_1,\dots,\bm q_n) = \bigintsss \left[\prod_{i=1}^n d^3y_i\,e^{-iq_{ia}y_i^a}\right]\,\delta^{(3)}\left(\frac{1}{n}\sum_{i=1}^ny_i\right)\Big\{u\cdot\partial G^{\rm c}_{i_1\dots i_n}\nn
  &\quad-
    n \left[\left(u^\mu\bar\omega^\nu_{\mu\lambda} y_1^\lambda
    \partial_\nu^{(y_1)} \delta_{i_1}^{\jj } + y_1^\mu\bar\omega^{\jj
    }_{\mu i_1}u\cdot\partial + u^\mu\bar\omega_{\mu i_1}^{\jj
    }\right)G^{\rm c}_{\jj i_2\dots i_n}\right]_{\overline{1\dots n}}
    \Big\}=\mathcal P[\{W_2,\dots, W_n\}]\,,
\end{align}
where in the last step one needs to perform the inverse Wigner transform of Eq.~\eqref{eq:W-ya} in order to obtain a set of local evolution equations for $W_n$, whose explicit form depends on Eq.~\eqref{eq:dtGn_c_general} that is
to be substituted into Eq.~\eqref{eq:cfd-W_c}.

\subsection{Rotating-wave approximation}
\label{sec:RPA}

Although the evolution equations appear diagrammatically simple as in Fig.~\ref{fig:diagrams-n}, the number of equations rapidly increases with the number of independent components of fluctuating fields, and the equations themselves also become more complicated. This complexity makes the numerical implementation challenging when considering additional hydrodynamic fields or higher-point functions. It is, therefore, instrumental to apply a method analogous to the rotating-wave approximation, under which the modes with rapid oscillation are averaged out. 

In this approximation, we first change fluctuation variables from the set we are using, say $\phi_i$, to a new set $\Phi_i$ defined in such a way that the linear operator $L$ appearing in the linearized ideal hydrodynamic equations is diagonal under linear transform $\phi = U\Phi$:
\begin{equation}
  \label{eq:lin-hydro-LL}
  u\cdot\partial\,\phi_i = L_{ij}\,\phi_j \quad\rightarrow\quad u\cdot\partial\,\Phi_i = \mathbb L_{ij}\Phi_j\,, \quad\mathbb L_{ij}=(U^{-1}LU)_{ij}=\lambda_i\delta_{ij}\,,
\end{equation}
where $\lambda_i$ are the eigenvalues of $L$, the transformation matrix $U$ can be constructed from the eigenvectors of the operator $L$, since for each index $j$ the vector with $i's$ component given by $U_{ij}$ is the {\em left\/} eigenvector of $L$: $\sum_iL_{ki}U_{ij}=\lambda_jU_{kj}$. The Wigner function transforms as $W_n\equiv W_{i_1\dots i_n}= U_{i_1j_1}\dots U_{i_nj_n}\mathbb W_{j_1\dots j_n}\equiv(U)^n\mathbb W_n$ accordingly, and Eq.~\eqref{eq:cfd-W_c} becomes
\begin{align}\label{eq:Wn-RWA}
  u\cdot\cfd \mathbb W_n&=\left((U)^nu\cdot\cfd (U^{-1})^n\right)\mathbb W_n+(U^{-1})^n\,\mathcal P[\{(U)^2 \mathbb W_2,\dots, (U)^n\mathbb W_n\}]=\left(\sum_{m=1}^n\lambda_{i_m}\right)\mathbb W_n+\dots
\end{align}
If $\lambda_{i_1}+\dots\lambda_{i_n}\neq0$, $\mathbb W_n\equiv\mathbb W_{i_1\dots i_n}$ is identified as a fast-oscillating mode that can be averaged out at a timescale much longer than the oscillation period. Consequently, we are left with only slow modes (with $\lambda_{i_1}+\dots\lambda_{i_n}=0$) whose evolution is described by relaxation-type equations.

\section{Summary}

In this proceeding we highlighted the essential ideas in the recent theoretical development of fluctuating hydrodynamics from the bottom-up approach. It consists of three ingredients: a closed system of equations for arbitrary $n$-point correlation functions, a formalism to describe these equations covariantly in relativistic hydrodynamics, and a strategy to reduce the number of correlation functions and simplify the resulting equations. The explicit full equations for the correlation functions that are relevant for heavy-ion collisions especially for the critical point search, as well as more detailed discussions will be presented in Ref.~\cite{An_2022b}.

\bibliographystyle{woc}
\bibliography{references}

\end{document}